\documentclass[12pt]{iopart}
\usepackage{amssymb}
\usepackage{graphics}
\begin{document}
\title{The ACIGA Data Analysis programme}
\author{Susan M Scott, Antony C Searle, Benedict J Cusack and David E McClelland}
\address{Department of Physics, Faculty of Science, The Australian National University, Canberra ACT 0200, AUSTRALIA}
\eads{\mailto{susan.scott@anu.edu.au}, \mailto{antony.searle@anu.edu.au}, \mailto{benedict.cusack@anu.edu.au},
\mailto{david.mcclelland@anu.edu.au}}
\begin{abstract}
The Data Analysis programme of the Australian Consortium for Interferometric Gravitational Astronomy (ACIGA) was set up in 1998 by the first author to complement the then existing ACIGA programmes working on suspension systems, lasers and optics, and detector configurations. The ACIGA Data Analysis programme continues to contribute significantly in the field; we present an overview of our activities.
\end{abstract}
\submitto{\CQG}
\maketitle
\section{Introduction}
ACIGA Data Analysis works in close collaboration with the Laser Interferometer Gravitational Wave Observatory (LIGO), making significant contributions to the LIGO Data Analysis System (LDAS) [1], in particular to its Data Conditioning component [2].  These systems underpin LIGO's forthcoming science results. We will describe the provision of spectral line removal code [3] and contributions of general signal processing operations.  The new ACIGA Data Analysis Cluster (ADAC) installed at The Australian National University (ANU) is an LDAS system used to characterise and validate our LDAS contributions and to analyse LIGO data and environmental data.  Some of this environmental data is locally acquired by a dedicated system operating at The ANU's Gravitational Wave Research Facility.  ACIGA is the only southern hemisphere participant in the international exchange of seismic, magnetic and grid voltage measurements, a process which marks the beginning of global co-operative gravitational wave astronomy [4]. We have simulated such global networks of observatories, drawing particular conclusions about the optimal placement of proposed new detectors, including a possible Western Australian observatory [5].

\section{The LIGO Data Analysis System (LDAS)}

LDAS is the data acquisition, storage, and analysis software and hardware system for the LIGO project.  ADAC is a fully-functional LDAS installation at The ANU in the Department of Physics.  It consists of three dual-processor servers respectively serving data, performing pipeline pre-processing, and managing a Beowulf cluster of eight single-processor nodes, with a terabyte of local storage and a link to the Mass Data Storage System (MDSS) at The ANU Supercomputer Facility (ANUSF).  It will give the team local capability to perform, on the extended datasets, the intensive characterisation required to fully validate complex algorithms like line removal, and to employ the algorithms of LDAS for independent ACIGA research on data from the LIGO observatories.  The cluster has already been used in the characterisation of the performance of line removal tools on actual LIGO data [3].

The Data Conditioning API (Application Programming Interface) is the component that performs the pre-processing of data in LDAS. It is a user-specified series of signal-processing `actions' on data flowing between other LDAS components. It is the component where all generic science tasks occur. ACIGA has contributed substantially to the design, development and testing of much of the Data Conditioning API, especially fundamental actions such as Discrete Fourier Transforms (DFTs), linear filtering, resampling, heterodyning, etc.\ and advanced algorithms using system identification techniques to remove artefacts.

One of the most visible instrumental artefacts in the output of interferometric observatories is the series of narrow spectral lines at 60 Hz and its harmonics due to the alternating-current power supply to the instrument. Unlike most instrumental noise, its cause can be monitored by sampling the supply voltage.  The line removal code developed at The ANU ingests these Physical Environment Monitor (PEM) channels and uses system identification techniques [6] to fit an optimal linear model to the system of the detector and disturbance, resulting in a prediction of the line shape using the supply voltage channel and thus the ability to subtract it away from the interferometric output. Initial characterisation [3] of the algorithm has been positive, and it has been used by the stochastic background search code.

In late 2002 we directly participated in the first science analyses of data taken from LIGO's first observing period, the ``S1 Science Run''.  Our line removal actions were integrated into the stochastic background pipeline, and the impact of correlated spectral lines on the stochastic background search codes was assessed [3, 7]. The analysis culminated in the setting of an upper limit on the strength of a cosmological background of gravitational radiation [7].

\section{Physical Environment Monitoring}

An obvious advantage gained by comparing data from a global network of detectors is immunity from local noise sources. There still exists, however, the possibility of global noise sources coupling parasitic signals into multiple detectors, so an effort is currently underway to understand such noise sources and evaluate their effect on the global array of gravitational wave detectors. In particular, three physical environment variables are being targeted: seismic disturbances, fluctuations in the Earth's magnetic field, and fluctuations in alternating-current power supply voltages. These variables are being monitored continuously at detector sites (LIGO Hanford, LIGO Livingston, VIRGO, GEO previously) and now at The ANU, and various comparison studies have been completed to date.

ACIGA's environment monitoring equipment is located at The ANU's Gravitational Wave Research Facility: three seismometers (for three axes), a three-axis magnetometer, and a simple grid voltage scaling and filtering circuit. A multi-channel analog-to-digital converter (ADC) and LabView-based data-logging software continuously log and process the data. The ANU's Mass Data Storage System (MDSS) stores the generated data in LIGO frame format ready for collection, via the rsync protocol, from an external party.

The ACIGA data-logging equipment uses a GPS receiver to synchronise samples with the other sites: the receiver's 1 Pulse-Per-Second signal triggers the initial sample at the start of a data taking run, and a function generator (using the receiver's 10MHz signal as a timebase) produces the 2048Hz sampling trigger required by the ADC board. The function generator limits the accuracy of the timing such that the system will lose one sample roughly every 200 days; the system is manually reset on a regular basis to allow for this, while still maintaining a $> 99$\% duty cycle.

An automated script on LIGO's High Performance Storage System (HPSS) collects frame-formatted data from the MDSS shortly after it is generated, and merges it with data from the other contributors. The HPSS, which supplies the prototype Network Data Analysis System (NDAS), has provided a test bed for synchronising and merging data from globally separated sources, culminating in the recent merging on the system of data taken from the second LIGO observing period, the ``S2 Science Run'', from LIGO, GEO and TAMA.

Correlation studies using the magnetometer data generated in the exchange have commenced recently, with our attempt to characterise and sanity-check the available data from the relevant channels.  Initially, studies have used medium-term timescales of $10^2$~s to $10^4$~s, for which the interpretation of correlation analyses is well understood. Second order correlation integrals (specifically, the coherence function) were computed using LIGO, ACIGA and VIRGO data.  Weak correlation lines were observed in some cases, for example at 50Hz between ACIGA and VIRGO (the fundamental power grid oscillation frequency at both sites), but the nature of this correlation (causal or coincidental) is not yet clear.  Some strong correlations were seen, such as a 16Hz line and its harmonics, thought to be caused by regular GPS-synchronised electronic events in the data-logging equipment at the LIGO Hanford and Livingston sites [7].

In addition, we have conducted a review of bicoherence estimates using ACIGA and LIGO magnetometer data [4].  Self-bispectra and cross-bispectra estimates are the third-order equivalents of power spectra and cross-spectra, respectively, (bicoherence is the normalised version of bispectra), and are useful for detecting nonlinearities based on two-frequency mixing [8], within a channel or across two channels.  Our review has found that no fundamentally different information could be obtained from the magnetometer channels by calculating bicoherences that could not be found with the second order equivalents, and so there appears to be no significant nonlinear processes in play with respect to the magnetometer channels.

\section{Global Network Optimisation}

Ground-based gravitational wave astronomy will eventually be performed using a global network of co-operating observatories. The ANU team has developed simulations to determine---with respect to the detection and characterisation of binary inspiral events, galactic neutron star emissions and other sources---the optimal location to site a new gravitational wave detector to augment an existing network of detectors, for a variety of data analysis strategies.

A common infrastructure for resolving the responses of gravitational wave detectors to sources depending on their relative location, orientation and polarisations was developed.  Particular analysis techniques---such as coincident and coherent inspiral detection---were simulated for stochastic populations of sources, enabling the comparison of different physical sites around the globe.

It was determined [5] that the siting of the $n$th detector in a network can have a substantial impact, up to a factor of two for some configurations, on the sensitivity of the network as a whole, and demonstrated that the Western Australian site of the proposed Australian instrument---near antipodal to the LIGO detectors---is optimal for augmenting the existing network. We continue to expand our simulations to include more general networks and other gravitational wave sources.

\section*{Acknowledgements}
This work
was supported by The ANU Faculties Research Grant Scheme, an award under
the Merit Allocation Scheme on the National Facility of the Australian Partnership for
Advanced Computation and the Australian Government's Systemic Infrastructure Initiative.

\section*{References}

1. http://ldas-cit.ligo.caltech.edu 
\\
2. http://ldas-cit.ligo.caltech.edu/doc\_index/datacond\_api.html 
\\
3. A C Searle, S M Scott and D E McClelland, Spectral line removal in the LIGO Data Analysis System (LDAS) 2003 Class.\ Quantum Grav.\ \textbf{20} S721-S730 
\\
4. B Cusack, A Searle, S Scott and D McClelland, Global second and third order correlations in physical environment monitors, http://www.ligo.caltech.edu/docs/G/G030364-00.pdf,
unpublished.
\\
5. A C Searle, S M Scott and D E McClelland, Network sensitivity to geographical configuration 2002 Class.\ Quantum Grav.\ \textbf{19} 1465-1470 
\\
6. L Ljung, System Identification: Theory for the User (2nd Ed.) Prentice Hall (1999) 
\\
7. B Abbott \emph{et al}, Analysis of first LIGO science data for stochastic gravitational waves, \emph{in preparation} 
\\
8. V Chickarmane, Bilinear coupling investigations, http://www.ligo.caltech.edu/docs/G/G030089-00.pdf, unpublished.

\end{document}